\documentclass[aps,preprint,secnumarabic,nobalancelastpage,nofootinbibt]{revtex4}

\usepackage{graphics}      
\usepackage{graphicx}      
\usepackage{longtable}     
\usepackage{url}           
\usepackage{bm,color}            
\usepackage{amssymb, amsmath, enumerate, theorem, epsfig,subfigure,setspace}
\usepackage{multirow}

\begin{document}

\title{Polaritonic XY-Ising Machine}

\author{Kirill P. Kalinin${}^1$, Alberto Amo${}^2$, Jacqueline Bloch${}^3$, and Natalia G. Berloff${}^{1,4}$}

\email[correspondence address: ]{N.G.Berloff@damtp.cam.ac.uk}
\affiliation{${}^1$Department of Applied Mathematics and Theoretical Physics, University of Cambridge, Cambridge CB3 0WA, United Kingdom}
\affiliation{${}^2$Univ. Lille, CNRS, UMR 8523 - PhLAM - Physique des Lasers Atomes et Mol\'ecules, F-59000 Lille, France}
\affiliation{${}^3$Centre  de  Nanosciences  et  de  Nanotechnologies, CNRS, Universit\'e  Paris-Sud, Universit\'e Paris-Saclay,  C2N Marcoussis, F-91460 Marcoussis, France}
\affiliation{${}^4$Skolkovo Institute of Science and Technology,  Bolshoy Boulevard 30, build. 1 Moscow,121205 Russian Federation}

\date{\today}

\begin{abstract}{Gain-dissipative systems of various physical origin have recently shown the ability to act as analogue minimisers of hard combinatorial optimisation problems. Whether or not these proposals will lead to any advantage in performance over the classical computations depends on the ability to establish controllable couplings for sufficiently dense short- and long-range interactions between the spins. Here, we propose a polaritonic XY-Ising machine based on a network of geometrically isolated polariton condensates capable of minimising discrete and continuous spin Hamiltonians. We elucidate the performance of the proposed computing platform for two types of couplings: relative and absolute. 
The interactions between the network nodes might be controlled by redirecting the emission between the condensates or by sending the phase information between nodes using resonant excitation. We discuss the conditions under which the proposed machine leads to a pure polariton simulator with pre-programmed couplings or results in a hybrid classical polariton simulator. We argue that the proposed architecture for the remote coupling control offers an improvement over geometrically coupled condensates in both accuracy and stability as well as increases versatility, range and connectivity of spin Hamiltonians that can be simulated with polariton networks.}
\end{abstract}

\maketitle

\textbf{Introduction}
Various physical systems have recently emerged as  unconventional computing platforms
with a potential to successfully compete against the established state-of-the-art classical techniques in solving large-scale combinatorial optimisation problems. Among the newly proposed computational machines are the coherent Ising machine (CIM) based on the degenerate optical parametric oscillators \cite{CIM2016}, a CMOS-based Fujitsu digital annealer \cite{Fujitsu_Helmut2019}, an all-electronic oscillator-based Ising machine \cite{ElectronicOsc_Ising2019}, a simulated bifurcation Toshiba machine \cite{Toshiba2019}, a spatial light modulator based photonic Ising machine \cite{PhotonicSLM2019}, and a molecular computing machine on a programmable microdroplet array \cite{Guzik_MolecularComputer2019}. All these approaches aim to achieve a much faster, more efficient and more accurate way of solving a particular class of optimisation problems, namely the quadratic unconstrained binary optimisation (QUBO) problem. In statistical physics QUBO problems appear when one seeks the global minimum (the ground state) of the Ising Hamiltonian. Indeed, the explicit polynomial mappings into the Ising Hamiltonian of many practically important problems of the discrete optimisation are available \cite{Lucas2014}, such as the travelling salesman, graph colouring,  warehouse inventory management, and low-risk portfolio optimisation. All these problems belong to the NP-hard computational complexity class meaning an exponential asymptotic growth of the number of operations with number of variables. It is no wonder that the advent of unconventional ways of finding the ground state of  the Ising Hamiltonian was accompanied by the development of new classical algorithms as well as new methods to compare the performance of different platforms. The former resulted in newly proposed physically-inspired algorithms \cite{KalininSciRep2018, Leleu2019} while the latter facilitated the design of instances of interaction matrices with the controlled complexity and planted solutions \cite{HelmutWishart2019,Lanl2019}. In addition to the Ising model, other spin Hamiltonians have been proposed for solving with physical platforms including the XY spin Hamiltonian simulators based on the photon condensates \cite{KlaersXYphoton2020}, coupled laser \cite{DavidsonXYlasers2019} and nanolaser arrays \citep{MarandiXYnanolasers2019}.

A system of polariton condensates has attracted a considerable interest over the last few years by offering an alternative gain-dissipative system for tackling both discrete and continuous optimisation problems. Polaritons are the bosonic quasi-particles arising from strong coupling between excitons and photons in a semiconductor microcavity \cite{bec}. Due to extremely low effective mass, polaritons can undergo Bose-Einstein condensation at temperatures higher than atomic condensates leading to two-dimensional networks of condensates operating at ambient conditions. Macroscopic coherence of such networks is  characterised by a complex classical field with a well-defined condensates' relative phases $\theta_i$. These phases can be mapped into  continuous `spins' ${\bf s}_i=(\cos \theta_i, \sin \theta_i)$ that can be further constrained to discrete values $\theta_i\in\{0,\pi\}$ by employing the resonant excitation \cite{GD-resonant}.  The idea underlying the polariton simulator for solving optimisation problems originates from the belief that a huge combinatorial space of possible states can be sought in parallel near the condensation threshold, at which only the low-energy coherent states can form. At a coherent state, the phase-locked condensates have the same frequency, chemical potential, and constant relative phases. This state may correspond to the local or global minimum of a particular spin Hamiltonian, and since condensation occurs on a picosecond time scale, such polariton simulators may be potentially attractive for optimisation tasks. 

Arbitrary networks of polariton condensates can be experimentally created in many different ways including optical imprinting \cite{wertz2010spontaneous,manni2011spontaneous,tosi2012sculpting,tosi2012geometrically},  in etched micropillars \cite{Bajoni2008,meanfield9,boulier2014polariton},
in lead halide perovskite lattices  \cite{su2018room,su2019observation}, in strain-induced traps \cite{balili2007bose}, in hybrid air gap microcavities \cite{dufferwiel2014strong}, by periodically etching the sample surface depositing metallic patterns \cite{Zhang_sample,Mischok_sample,Lai2007}, by surface acoustic waves \cite{cerda2012dynamic}, by direct fabrication with the gold deposition technique \cite{kim2011dynamical}, in coupled mesas etching during the growth of the microcavity \cite{winkler2015polariton}. The first of these, optical imprinting, is commonly realised with a liquid crystal spatial light modulator (SLM). The robustness of this technique has been demonstrated for a variety of lattice configurations and sizes \cite{wertz2010spontaneous,manni2011spontaneous,BerloffNatMat2017,OhadiTrapped2017,OhadiTrapped2018} proving the scalability of the polariton system for both trapped geometries and freely expanding condensates. The coupling between geometrically coupled condensates is generally of a complex nature \cite{OurPRB2019} and consists of the dissipative (Heisenberg) and Josephson couplings. The latter prevents the system from achieving the minimum of a spin Hamiltonian as was previously elucidated \cite{OurPRB2019}. Even when the Josephson coupling is negligible in comparison with the dissipative coupling, the geometric coupling barely allows one to control the interactions beyond the nearest neighbours. This prevents the system from addressing complex, non-planar spin Hamiltonians. 
Nevertheless, the mapping of combinatorial problems into polariton networks was originally suggested by controlling the distance between geometrically coupled condensates \cite{BerloffNatMat2017,ourDM} and later followed by propositions to control interactions in regular lattices of condensates with dissipative gates \cite{ourAdvTechn2019} or resonant pump barriers \cite{PavlosBarriers}. The initial experimental demonstration of minimising the XY spin Hamiltonian with simple polarion lattice cells \cite{BerloffNatMat2017} has been shortly followed by a theoretical proposal extending the class of simulated spin Hamitonians from continuous to discrete \cite{GD-resonant}.

Finding ways to dynamically control individual interactions between network nodes is a necessary step for addressing non-trivial spin Hamiltonians but not sufficient. In all proposed schemes the nearest neighbour interactions are attempted to be controlled while the beyond nearest neighbour interactions are assumed to be negligible, which is rarely the case. A recent study has shown the synchonisation between condensates across distances over 100 $\mu m$  \cite{PavlosTimeDelay} noting that a typical lattice size constant is often in the range of 5-15$\mu m$ \cite{BerloffNatMat2017}. This leads to a crucial and yet missing discussion of controlling the couplings beyond nearest neighbours for arbitrary graphs of polariton condensates. Moreover, spatially coupled polariton condensates are capable of representing different oscillator models \cite{OurPRB2019} including the Kuramato, Sakaguchi-Kuramoto, Lang-Kobayashi and Stuart-Landau models for different ranges of experimental parameters some of which are easier to adjust, e.g. exciton-polariton interactions, while others are harder, e.g. polariton lifetime. This apparent flexibility of a polariton system makes it harder to isolate a particular optimisation problem to address with polariton networks and, consequently, limits the optimisation accuracy of any objective function. 
To distinguish between many models that can be modelled with polariton networks, an instrumental calibration of experimental parameters may be required for spatially coupled polariton condensates even for nearest neighbour interactions. 

In this work, we offer an alternative approach for simulating spin Hamiltonians with a network of spatially localised polariton condensates that do not interact with one another geometrically. For the network to become a spin Hamiltonian optimiser, we propose to couple any two condensates  by redirecting the emission from one condensate to another or by exciting one condensate with an additional resonant pump tuned to the phase of that condensate. We emulate the polariton simulator with the two-dimensional complex Ginzburg-Landau equation coupled to the exciton reservoir equation and consider two possible coupling schemes, namely {\it relative} and \textit{absolute} couplings. The performance of the emulated polariton simulator is demonstrated for discrete, i.e. Ising, and continuous, i.e. XY, spin Hamiltonians for sparse and dense interaction matrices $J$ of various sizes from 9 to 49 condensates. This manuscript does not introduce a new optimisation algorithm and, therefore, the common metrics for comparing algorithms, e.g. time-to-solution, are intentionally omitted. Instead, we offer a proof-of-concept of a real XY-Ising polariton machine that can be realised within the vast range of experimental parameters. We outline the conditions under which the proposed polariton simulator becomes a ``pure" or ``hybrid-classical" physical optimiser.

%

\textbf{Results}
 
Over the last decade, polariton condensates have been successfully modelled \cite{goveq, carusotto,reviewKeeling,reviewCarusotto}  by the generalised complex Ginzburg-Landau equation (cGLE) coupled to the exciton reservoir dynamics
\begin{eqnarray}
	i \hbar  \frac{\partial \psi}{\partial t} &=& - \frac{\hbar^2}{2 m}(1 - i \eta {\cal R})  \nabla^2\psi + U_0 |\psi|^2 \psi + g_R {\cal R} \psi 	  + \nonumber\\
	& +& \frac{i \hbar}{2} [R_R {\cal R} - \gamma_C] \psi +
	i f_{\rm Res}({\bf r},t) \psi^*, \\
	\frac{\partial {\cal R}}{\partial t} &=&  - \left(\gamma_R+ R_R|\psi|^2 \right) {\cal R} + P({\bf r},t),
\end{eqnarray}
where $\psi({\bf r},t)$ is the condensate wavefunction, $m$ is the effective polariton mass, ${\cal R}({\bf r},t)$ is the density distribution of the exciton reservoir, $U_0$ and $g_R$ are the polariton-polariton and polariton-exciton  interaction strengths respectively, $R_R$ is the rate of scattering from the reservoir into the condensate, $\eta$ is the energy relaxation. 
The $f_{\rm Res}$ term is an optional resonant pump at the double condensate frequency (second resonance) which forces phase differences between different condensates to be either 0 or $\pi$ \cite{GD-resonant}. 
Linear photon losses in the cavity and exciton losses are described by the condensate ($\gamma_C$) and the reservoir ($\gamma_R$) damping rates.   The pumping intensity $P({\bf r},t)$ characterises the incoherent excitation. The process of Bose-Einstein condensation is essentially quantum, however,  once a condensate is formed it can be accurately described by the cGLE as was shown in numerous experimental works \cite{tosi2012sculpting,meanfield2,tosi2012geometrically,meanfield4,meanfield5,meanfield6,meanfield7,meanfield8,meanfield9}.  
The optimisation accuracy of either XY or Ising models does not change for small values of $\eta$ that are asumed in experiments \cite{GD-resonant} and, therefore, we will neglect it for the rest of the manuscript with a note that the bigger values would have had a negative effect on the accuracy. Introducing the dimensionless variables and parameters   by $\psi \rightarrow \sqrt{ \hbar R_R / 2 U_0 l_0^2} \psi$, $t \rightarrow 2 l_0^2 t / R_R$, $ {\bf r} \rightarrow \sqrt{\hbar l_0^2 / (m R_R)} {\bf r}$, ${\cal R} \rightarrow {\cal R} / l_0^2$, $P \rightarrow R_R P / 2 l_0^2$, $f_{\rm Res} \rightarrow 2 l_0^2 f_{\rm Res}/ \hbar R_R$, $g= 2 g_R / R_R$, $b_0= 2 \gamma_R l_0^2 / R_R$, $b_1=  \hbar R_R / U_0$, $\gamma =  \gamma_C l_0^2 / R_R$, $l_0 = 1 \mu m$, we obtain
 \begin{eqnarray}	
	i  \frac{\partial \psi}{\partial t} &=& -  \nabla^2\psi +  |\psi|^2 \psi+
g {\cal R} \psi   + 	\nonumber\\
		&+& i({\cal R} - \gamma) \psi +  i f_{\rm Res}({\bf r},t) \psi^*, \label{e1}\\
	  \frac{\partial {\cal R}}{\partial t} &=&  - \left(b_0+ b_1|\psi|^2 \right) {\cal R} + P({\bf r},t).
	\label{e2}
\end{eqnarray}

The cGLE (Eq.~(\ref{e1})) is a universal order parameter equation that describes non-linear phenomena in driven-dissipative systems ranging from non-linear waves and second-order phase transitions to lasers and superconductors. In this work, we use these equations to represent a network of isolated non-interacting polariton condensates which can be experimentally realised, for instance, with micropillars or with trapped polariton condensates. Although the following analysis can be readily applied to either experimental configuration, for ease of reading we will use an array of micropillars as our primary example of isolated condensates with occasional notes on the possible change in  performance of the other. The position, shape and size of each micropillar can be accurately controlled during their fabrication  \cite{boulier2014polariton}. Hundreds of coupled micropillars etched in a planar semiconductor microcavity have been used to study a wealth of phenomena from the Dirac cones in a honeycomb geometry \cite{jacqmin2014direct} to the gap solitons in 1D Lieb lattices \cite{goblot2019nonlinear}. To model the polariton condensation in a micropillar cavity, we introduce a spatially dependent dissipative profile
\begin{equation}
\gamma({\bf r})= \gamma_{\rm out} - (\gamma_{\rm out} - \gamma_{\rm in}) \sum_i \exp\big( - \alpha | {\bf r} - {\bf r}_i |^{2 n_{\rm SG}} \big),
\end{equation}
where $\gamma_{\rm out}$ and $\gamma_{\rm in}$ are the dissipation rates outside and inside of a micropillar, respectively. Here, $\gamma_{\rm out}\gg\gamma_{\rm in},$   ${\bf r}_{i}$ denotes the center of the $i$-th micropillar, and $n_{\rm SG}$ is the degree of a supergaussian that models micropillars as flat low-dissipative discs. The dramatically increased dissipation between the discs  ($\gamma_{\rm  out} = 100 \gamma_{\rm in}$) effectively blocks all the polariton outflows which leads to non-interacting condensates even for short separation distances of a few micrometers as would be expected for the system of micropillars. The condensates at different micropillars are noninteracting unless   either {\it relative} or {\it absolute} remote couplings are introduced. In the former case, a part of the light emitted by the  $j$-th micropillar condensate is re-injected into the $i$-th micropillar condensate at the amount proportional to the occupation of the $j$-th condensate. In the case of the {\it absolute coupling}, the same amount of light is exchanged between the $i$-th and the $j$-th condensates.
Both coupling models can be represented by
\begin{eqnarray}	
	i   \psi_t &=& -  \nabla^2\psi +  |\psi|^2 \psi+
g {\cal R} \psi	 + i({\cal R} - \gamma) \psi +i f_{\rm Res} \psi^*\nonumber \\
&  +&   i \delta_{\gamma, \gamma_{\rm in}} \sum_{j=1,j\ne i}^N {\cal J}_{ij} \psi({\bf r + r}_j - {\bf r}_i, t- \tau) \label{relative}
\end{eqnarray}
where $\delta_{\gamma, \gamma_{\rm in}}$ is the delta-function which is equal to one inside a micropillar, i.e when $\gamma({\bf r}) = \gamma_{\rm in}$, and zero outside, $N$ is the number of micropillars, and $\tau$ represents a possible time-delay to supply couplings in an experimental setup. The coupling term represents the emission feedback when for each $\psi({\bf r})$ in a micropillar $i$ the respective values $\psi({\bf r + r}_j - {\bf r}_i)$ are added from the  micropillar centred at $j$. For the \textit{relative} coupling model we shall consider ${\cal J}_{ij} = J_{ij}$ while for the \textit{absolute} coupling model we will use ${\cal J}_{ij} = |\psi_i| J_{ij} / |\psi_j|$. The sign of the coupling strength can be made positive or negative by injecting the light with zero (ferromagnetic coupling) or $\pi$ phase (anti-ferromagnetic coupling), respectively. For further derivations, we denote $K(r)=\Theta(R-r)$ as the Heaviside function where $R$ is the radius of the central part of the micropillar with a uniform phase distribution. In Eq.~(\ref{relative}) we assume that the frequencies of each individual micropillar may be slightly different just below the condensation threshold. Nevertheless, the condensation process in presence of interpillar couplings locks these frequencies of different condensates resulting in a single energy condensate level~\citep{OurNJP2018}. Recent experimental reports on two coupled micropillar lasers have demonstrated such frequency locking for detunings of up to 1GHz in the few photons regime~\cite{Schlottmann2016, Kreinberg2019}. For negligible time-delay and geometric couplings between condensates, we can rewrite Eqs.~(\ref{e2},\ref{relative}) for each micropillar $i$ using $\psi=\sum_i \psi_i$ and ${\cal R} = \sum_i {\cal R}_i$ as $N$ equations for the polariton condensates $\psi_i=K(|{\bf r-r}_i|)\psi({\bf r-r}_i, t)$  and $N$ equations for reservoir densities ${\cal R}_i = K(|{\bf r-r}_i|) {\cal R}({\bf r-r}_i, t)$ noting that $P({\bf r}, t) = \sum_i P_i = \sum_i P(|{\bf r - r}_i|, t)$, $f_{\rm Res}({\bf r}, t) = \sum_i f_{\rm Res}^{(i)} = \sum_i f_{\rm Res}(|{\bf r - r}_i|, t)$:
\begin{eqnarray}	
	i   \partial_t \psi_i&=& - \nabla^2\psi_i +  |\psi_i|^2 \psi_i+
g {\cal R}_i \psi_i +  i({\cal R}_i  - \gamma_{in}) \psi_i 	+  i f^{(i)}_{\rm Res} \psi_i^* + i \sum_{j=1,j\ne i}^N {\cal J}_{ij} \psi_j \label{relative-rates}\\
\partial_t {\cal R}_i &=&  - \left(b_0+ b_1|\psi_i|^2 \right) {\cal R}_i + P_i.
	\label{nri}
\end{eqnarray}
The steady states of  Eqs.~(\ref{relative-rates}-\ref{nri}) correspond to the minima of the XY, i.e. $f^{(i)}_{\rm Res} = 0$, and Ising, i.e. $f^{(i)}_{\rm Res} \neq 0$, models as it becomes evident after we substitute $\psi_i = \sqrt{\rho_i} \exp[i \theta_i]$ into Eq.~(\ref{nri}) and separate the real and imaginary parts. The equations read as
\begin{eqnarray}	
	\frac{1}{2}\partial_t \rho_i &=&  ({\cal R}_i  - \gamma_{in}) \rho_i + \sum_{j=1,j\ne i}^N \sqrt{\rho_i \rho_j} {\cal J}_{ij} \cos(\theta_{ji})  +  \rho_i f_{\rm Res}({\bf r},t) \cos(2\theta_i), \label{rho_t}\\
\partial_t \theta_i &=& \frac{\nabla^2 \sqrt{\rho_i}}{\sqrt{\rho_i}} - \rho_i - g {\cal R}_i + \sum_{j=1,j\ne i}^N \sqrt{\frac{\rho_j}{\rho_i}} {\cal J}_{ij} \sin(\theta_{ji})  -  f_{\rm Res}({\bf r},t) \sin(2\theta_i), \label{theta_t}\\
\partial_t {\cal R}_i &=&  - \left(b_0+ b_1 \rho_i \right) {\cal R}_i + P_i(|{\bf r-r}_i|,t), \label{R_t}
\end{eqnarray}%
where $\theta_{ji} = \theta_j - \theta_i$. Here we considered the uniform phase distribution $\theta_i(|{\bf r - r}_i|, t) \approx \theta_i(t)$  which is a valid assumption near the micropillar's centre, i.e. $R < R_{m}$ with $R_m$ being the micropillar's radius. In case of the {\it relative} coupling scheme, the fixed points of Eqs.~(\ref{rho_t}-\ref{R_t}) represent the minima of the XY or Ising spin Hamiltonians only for the equal polariton densities \cite{OurNJP2018} across all micropillars, that is, when the condition $\rho_i({\bf r}) = \rho_j({\bf r})$ stands. Such density equilibration can be robustly achieved by iteratively updating pumping intensity $P_i$ so that  $\int \rho_i d{\bf r} = \rho_{\rm th}$ for all micropillars, where $\rho_{\rm th}$ is the predefined integral luminosity. In contrast, the \textit{absolute} coupling model naturally optimises the XY and Ising models and doesn't require the equalised polariton densities due to the coupling coefficients ${\cal J}_{ij} = |\psi_i| J_{ij} / |\psi_j|$ that represent the exchange of a fixed number of photons between sites. The steady state solution of Eqs.~(\ref{rho_t}-\ref{R_t}) is given for both coupling models by equations:
\begin{eqnarray}	
& & ({\cal R}_i  - \gamma_{in}) = - \sum_{j=1,j\ne i}^N J_{ij} \cos(\theta_{ji})  -f_{\rm Res}^{(i)} \cos(2\theta_i), \nonumber \label{eq_fixed_point}\\ 
\\
& & \mu -\frac{\nabla^2 \sqrt{\rho_i}}{\sqrt{\rho_i}} + \rho_i + g {\cal R}_i = \sum_{j=1,j\ne i}^N  J_{ij} \sin(\theta_{ji})  - f_{\rm Res}^{(i)}\sin(2\theta_i) \nonumber \\
\\
& &{\cal R}_i  = P_i(b_0+ b_1 \rho_i)^{-1}, \label{Ri_fixed_point}
\end{eqnarray}
where $\mu$ is the global oscillation frequency shared between all condensates at a coherent state. 

One can see from the Eq.~(\ref{Ri_fixed_point}) that for a fixed point solution the maximised total polariton density corresponds to the minimum of the total reservoir density, which together with Eq.~(\ref{eq_fixed_point}) leads to the minimisation of the spin Hamiltonians:
\begin{eqnarray}	
& & \max \sum_{i = 1}^N \int_\Omega \rho_i d{\bf r} \Leftrightarrow \min \sum_{i = 1}^N  \int_\Omega {\cal R}_i d{\bf r} \Leftrightarrow \min H_{\rm XY|Ising} \nonumber \\
& & H_{\rm XY|Ising} = - \frac{1}{2}\sum_{i,j = 1}^N J_{ij} \cos(\theta_{ij})  -\sum_{i = 1}^N \big( \int_\Omega f_{\rm Res}^{(i)} d{\bf r} \big) \cos(2\theta_i), \nonumber
\end{eqnarray}
where $\Omega$ denotes the plane of the microcavity. The resonant force term $f^{(i)}_{\rm Res}$ acts as a penalty in the objective function and leads to optimisation of the Ising model while the XY Hamiltonian is optimised for zero penalty term. We note that the term $g {\cal R}_i$ has a destabilising effect on the steady states solutions corresponding to minima of spin Hamiltonians meaning that small exciton-polariton interactions and/or small exciton reservoirs ${\cal R}_i$ could possibly improve the optimisation accuracy. In experiments, a small reservoir density can be achieved for a high conversion rate of excitons into polaritons or by spatially separating polaritons from the reservoir by considering, for example, trapped condensates.

\begin{figure}[h!]
	\centering
	\includegraphics[width=8.6cm]{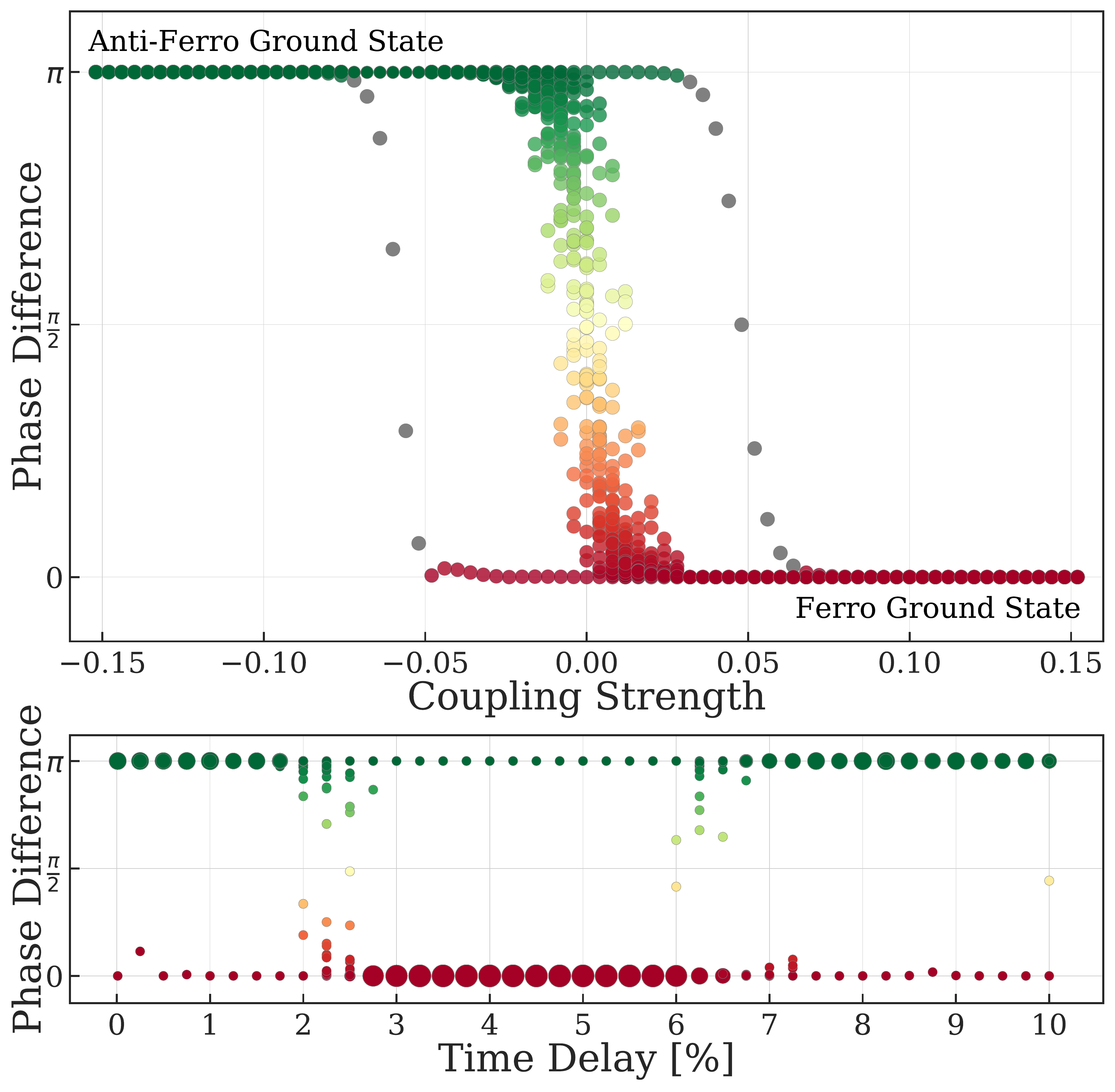}
	\caption{Top: Phase difference as a function of coupling strength for a polariton dyad. The Eqs.~(\ref{relative-rates}-\ref{nri}) are simulated for 50 random initial conditions for each coupling value. The coherence occurs for the absolute values of strengths greater than $0.02$ leading to ferromagnetic state with $0$ phase difference for positive couplings and to antiferromagnetic state with $\pi$ phase difference for negative couplings. The slowly decaying unstable solutions are shown in grey. Bottom: Phase difference as a function of time-delay for a polariton dyad. The time-delay percentage is defined with respect to the time required to reach a steady state in the absence of delay. The scatter point size indicates how many states out of 50 initial conditions end with a particular phase difference.  The coupling strength between condensates is chosen to be $J_0 = -0.1$. The expected anti-ferromagnetic state is observed for time-delays $\tau < 2\%$ and followed by the region with decoupled condensates. The further phase-locking of condensates becomes possible for bigger time-delays due to the global phase presence.}
    \label{Fig1}
\end{figure}
The validity of the proposed \textit{relative} and \textit{absolute} coupling models is verified by applying the two-dimensional Eqs.~(\ref{relative-rates}-\ref{nri}) for optimisation of the XY and Ising Hamiltonians on various coupling matrices. Firstly, we determine the minimum value of the coupling strength required for phase-locking of two condensates. Figure~\ref{Fig1}(top) shows the phase difference for a polariton dyad in the case of different interaction strengths with a zero time-delay. For each coupling strength, we simulate 50 random initial conditions and calculate the phase difference between the condensates in a final steady state. The region of decoupled condensates can be identified for coupling strengths $|J_0| \lesssim 0.02$ by observing random phase differences between the condensates in Fig.~\ref{Fig1}(top). For bigger coupling strengths, the condensates become phase locked and can reach ferromagnetic  ground state (with zero phase difference between the condensates) for the positive couplings or antiferromagnetic ground state (with $\pi$ phase difference) for the negative couplings. The local minima become unstable for coupling strengths bigger than $0.05$ and the system finds the ground state regardless of the initial conditions. The demonstrated minimum coupling strengths for phase-locking of two condensates are similar for both {\it relative} and \textit{absolute} coupling models in case of the XY Hamiltonian. For the Ising Hamiltonian, the destabilisation of excited states (local minima) happens for bigger coupling strengths of about $|J_0| \geq 0.07$. This is therefore the minimum coupling strength needed for the system to find the dyad's ground state independently of the initial conditions. We note that the presence of intrinsic noise has a positive effect on destabilising such local minima.

In an experimental implementation of interactions, a possible time-delay $\tau$ may appear in constructing couplings between the network elements due to multiple reasons including the phase readout time, the time required to re-route photons, or the time for adjusting an SLM. As a result, the delayed phase information of condensates at time $t - \tau$ will be used for creating couplings between the condensates at time $t$ whose phases will be shifted due to the global oscillation frequency. To demonstrate this effect of a time-delay in realising coupling strengths between different micropillars, we consider the \textit{absolute} coupling model in optimising the XY Hamiltonian. Figure~\ref{Fig1}(bottom) shows the phase difference dependence on the time-delay for the polariton dyad with the coupling strength $J_0 = -0.1$. The percentage time-delay is defined as a ratio to the time $T$ that is required for the dyad to reach a steady state in the absence of the delay. For each time-delay value, we simulate 50 random initial conditions and show the resulting phase difference with scatter points of varied sizes proportional to the fraction of initial conditions that lead to this phase. The anticipated anti-ferromagnetic ground state is observed for time-delays $\tau$ up to $2\%$. The previously unstable local minimum, i.e. ferromagnetic state with $0$ phase difference for $J_0 = -0.1$, becomes now stable in the presence of time-delay. Interestingly, the subsequent de-synchronisation area is followed by a clear ferromagnetic coupling between condensates which is in turn followed by another anti-ferromagnetic area for $\tau > 7\%$. This peculiar synchronization behaviour can be attributed to the global phase rotation with frequency $\mu$ of each condensate which can lead to phase-locking of condensates with an additional $\pi$ phase difference for large time-delay values. This time-delay effect is similar for both coupling schemes in simulating either spin Hamiltonian.  Although for networks of condensates, the presence of a time-delay would result in a phase lag \cite{OurPRB2019}  in Eqs.~(\ref{Ri_fixed_point}-\ref{eq_fixed_point}) which for significant $\tau$ can decrease the optimisation accuracy of the XY Hamiltonian, but not Ising. For simplicity, in the following investigations, we will not consider any time-delay in the couplings.

\begin{figure}[b!]
	\centering
	\includegraphics[width=8.6cm]{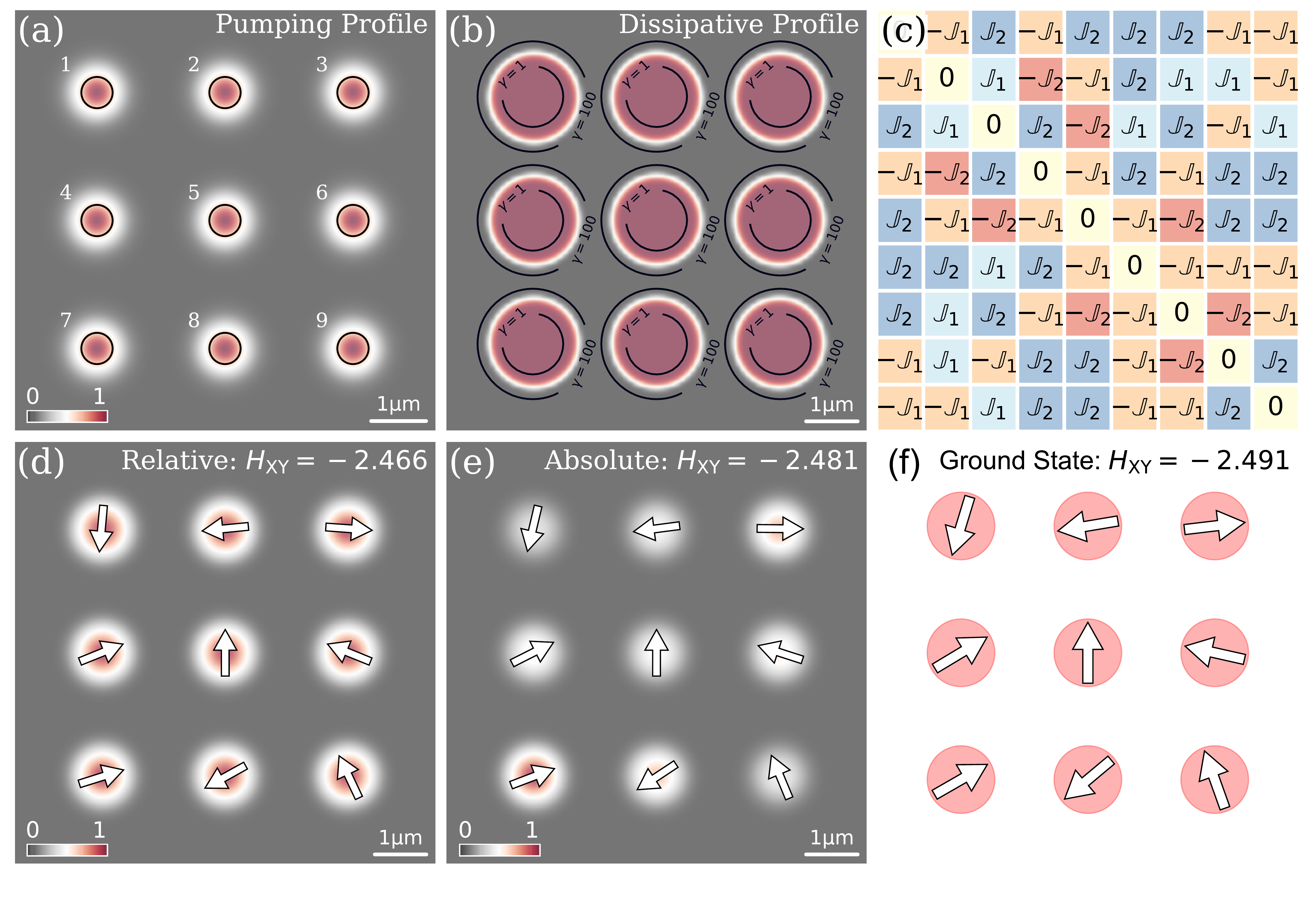}
	\caption{Finding the global minimum of  the XY Hamiltonian of size $N = 9$ with a $3 \times 3$ polariton lattice by simulating  Eqs.~(\ref{relative-rates}-\ref{nri}). (a) The intensity distribution of the  incoherent pumping profile $P({\bf r})$. The condensate emissions within the black circles are used for  couplings between condensates. (b) The dissipative profile for realising spatially isolated polariton pillars. (c) The fully-connected coupling matrix $J$ which is  randomly constructed from $J_1 = 0.05$ and $J_2 = 0.1$ of random signs. (d-e) The polariton density profiles  and phase configurations are plotted for {\it relative} and \textit{absolute} coupling models, respectively. The white arrows represent the phase difference with respect to the central condensate (with respect to the vertical arrow). The corresponding energy values of the $XY$ Hamiltonian are shown in the top-right corner. (f) The ground state solution of the XY Hamiltonian is verified by the gain-dissipative and the basin-hopping algorithms.}
    \label{Fig2}
\end{figure}

Having established the minimum coupling strength for phase-locking of two condensates, we now 
consider nine fully-connected polariton condensates. Each condensate is created with a non-resonant Gaussian pump in a lattice of 3 by 3 condensates (see Fig.~\ref{Fig2}(a)). To realise  spatially non-interacting polariton condensates we introduce  a dissipative profile as shown in Fig.~\ref{Fig2}(b) where the absence of particle outflows is ensured by the high value of $\gamma_{\rm out} = 100$ outside nodes compared to low $\gamma_{\rm in} = 1$ values inside nodes. 
 A random interaction matrix is constructed of positive and negative couplings of amplitude $\{0.05, 0.1\}$ as shown in Fig.~\ref{Fig2}(c). As an illustrative example, we apply the {\it relative} and \textit{absolute} coupling models described by Eqs.~(\ref{relative-rates}-\ref{nri}) for optimising the XY Hamiltonian ($f_{\rm Res} = 0$).
In the former case, the densities of condensates are iteratively equalised over time by individually adjusting pumping intensities $P_i$. The \textit{absolute} coupling model does not require equal polariton densities at the steady state and, consequently, non-equal densities can be realised in a final state. The phase configurations and corresponding density profiles are shown in Fig.~\ref{Fig2}(d-e) for the lowest energy states out of 10 runs for both models. To quantify the optimisation performance of coupling models, we consider the median accuracy that is defined by a proximity to the ground state:
\begin{equation}
Median \ Accuracy = < \frac{H_{\rm Relative|Absolute}}{H_{\rm Ground \ State}}>.
\end{equation}
where $H_{\rm Relative|Absolute}$ is the spin Hamiltonian energy for the phase configurations obtained with the mean-field approach (Eqs.~(\ref{relative-rates}-\ref{nri})) in case of the \textit{relative} or \textit{absolute} coupling schemes, $H_{\rm Ground \ State}$ is the ground state energy found by the classical optimisation algorithms.
In Fig.~\ref{Fig2}(d-e), the found minima are within $1\%$ and $0.4\%$ from the ground state  of the XY Hamiltonian  that was verified with the gain-dissipative \cite{KalininSciRep2018} and the basin-hopping \cite{BasinHopping} algorithms (Fig.~\ref{Fig2}(f)). The median accuracy over 100 random fully-connected matrices of size $N = 9$ generalises to $99.2\%$ and $99.5\%$ for the XY Hamiltonian in case of the \textit{relative} and \textit{absolute} coupling models, respectively. 

\begin{figure}[h!]
	\centering
	\includegraphics[width=8.6cm]{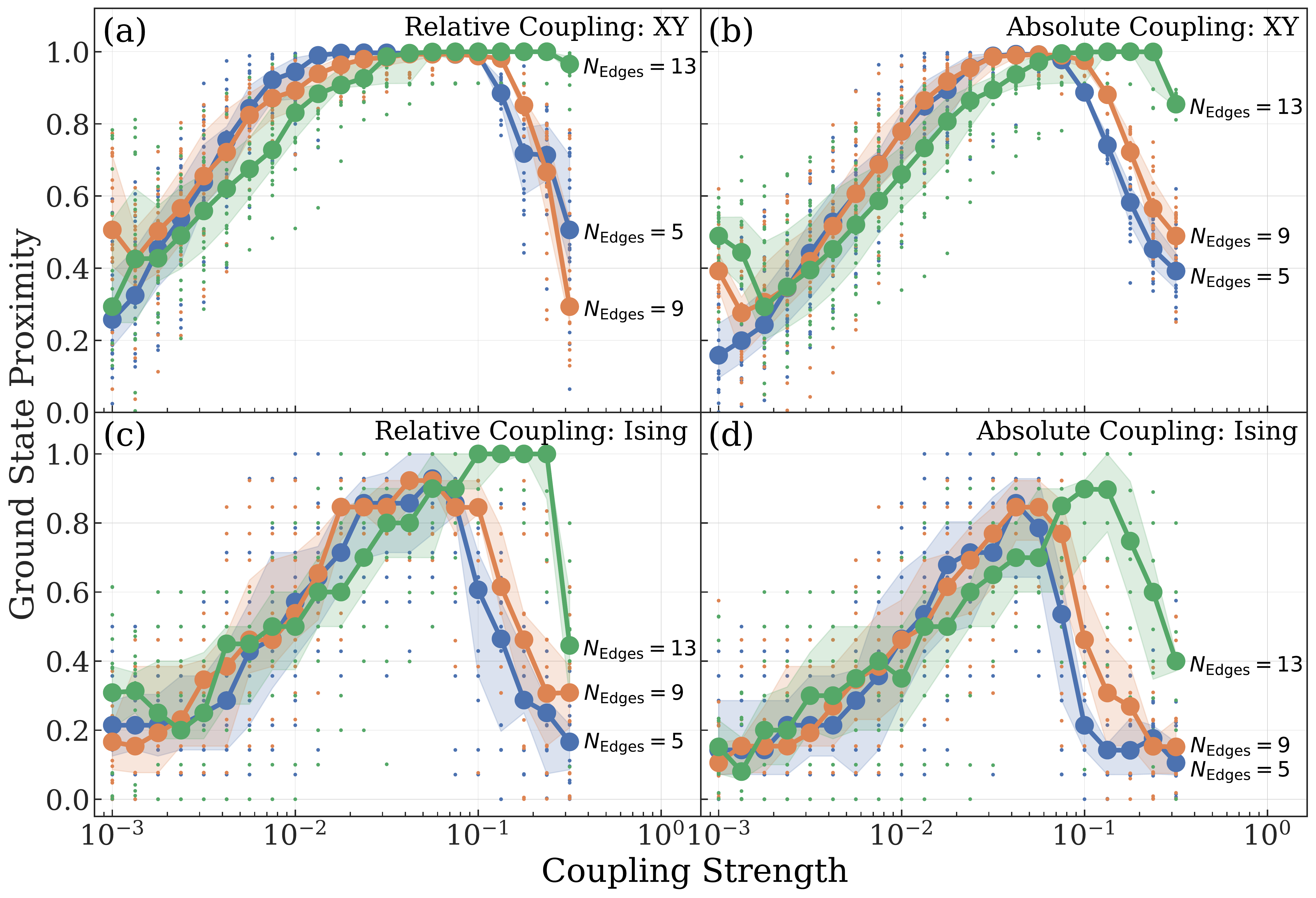}
	\caption{Optimal amplitude range study for {\it relative} and \textit{absolute} coupling models on the unweighted MaxCut problems of size $N = 16$ with degrees 5, 9, and 13. The median accuracy is shown for the XY Hamiltonian in (a-b) and the Ising Hamiltonian in (c-d). Both models are simulated with Eqs.~(\ref{relative-rates}-\ref{nri}) for 20 random initial conditions per each coupling strength. Shading indicates 25th and 75th percentile range of instances.}
    \label{Fig3}
\end{figure}

To investigate the performance of the proposed polaritonic XY-Ising machine on the bigger size problems, an analysis of the optimal range of coupling values and edge density effects is required. In what follows we study the {\it relative} and \textit{absolute} coupling models on the random unweighted MaxCut problems for the XY and Ising spin Hamiltonians. For the unweighted MAX-CUT problem, one seeks to divide the graph into two subgraphs with the maximised number of edges between them. This problem is known to be NP-hard \cite{UnweightedMaxCut} and can be mapped to the Ising Hamiltonian by assigning antiferromagnetic couplings $J_{ij} = -1$ to the graph edges. We construct three such random adjacency matrices ${\cal A}$ of size $N = 16$ of degree 5, 9, and 13. Both coupling models are simulated on matrices $J = -J_0 {\cal A}$ with amplitudes $J_0$ in the range $[0.001, 0.3]$. For each coupling strength amplitude, the Eqs.~(\ref{relative-rates}-\ref{nri}) are simulated for 20 random initial conditions. Figure~\ref{Fig3}(a-b) shows the ground state proximity as a function of $J_0$ amplitude for the XY Hamiltonian. The optimal range of couplings with the median accuracy over $90\%$ can be identified for the amplitudes in the range $[0.01, 0.11]$ for the {\it relative} coupling model and slightly smaller range of $[0.02, 0.1]$ for the \textit{absolute} coupling model. For the Ising model, a smaller batch of coupling amplitudes allows one to achieve the median accuracy greater than $90\%$ (see Fig.~\ref{Fig3}(c-d)). Such difference between the optimal coupling ranges can be possibly anticipated since the hard problems for the Ising Hamiltonian are not necessarily hard for the XY Hamiltonian optimisation. The clear shift to bigger optimal couplings for bigger edge densities ($>0.8$) is especially pronounced for the Ising Hamiltonian. This analysis confirms the lower bound and provides the upper bound of the coupling strength $J_0$ for achieving higher optimisation accuracies for both models.
We note that the  ground states of the Ising Hamiltonians were verified with the gain-dissipative \cite{KalininSciRep2018} and CIM \cite{Leleu2019} algorithms.

\begin{table}[b!]
\begin{ruledtabular}
\caption{Optimisation of the Ising and XY spin Hamiltonians with {\it relative} and \textit{absolute} coupling models on unweighted MaxCut problems of size 25 and 49 with edge density $0.5$. The median accuracy of both models is calculated for 20 random initial conditions per each coupling matrix which was further averaged over 20 random coupling matrices with coupling strength $J_0 = 0.04$. The number in parentheses indicates how many problems with different coupling matrices were globally optimised. The ground state solutions are calculated with the gain-dissipative and the basin-hopping algorithms for the XY Hamiltonians and the gain-dissipative and CIM algorithms for the Ising Hamiltonians.}
\centering
\label{table:performance}
\begin{tabular}{l c c c c}
\multirow{2}{*}{Problem Size} & \multicolumn{2}{c}{Relative} & \multicolumn{2}{c}{Absolute} \\ 
  & XY & Ising  & XY & Ising  \\ 
\hline
25 ($5 \times 5$ lattice) & 99.3\% (20)& 87.8\% (20)& 96.8\% (20) & 72.9\% (14)\\
49 ($7 \times 7$ lattice) & 98.2\% (20) & 81.7\% (4) & 93.3\% (3) & 52.3\% (0) \\
\end{tabular}
\end{ruledtabular}
\end{table}

With the identified optimal range of coupling amplitudes, we apply the {\it relative} and \textit{absolute} coupling models to bigger spin Hamiltonian problems. Table~\ref{table:performance} shows the median accuracy for both coupling models simulated on 20 unweighted MaxCut instances of size 25 and 49 with edge density of $50\%$. For such connectivity, we pick the amplitude strength of $J_0 = 0.04$ from the optimal range. The number of initial conditions is fixed to 20 per each coupling matrix. We say that the coupling matrix $J$ is globally optimised if the actual ground state is found at least once out of 20 runs for the Ising Hamiltonian. In case of the XY Hamiltonian, we require at least one phase configuration that is closer than $98\%$ to the ground state for claiming global optimisation. This number of globally optimised interaction matrices is indicated in parentheses in Table~\ref{table:performance}. The \textit{relative} coupling model shows a consistently better performance on both the Ising and XY Hamiltonians than the \textit{absolute} coupling model. The less accurate results for the Ising Hamiltonian, which are even more pronounced for the \textit{absolute} coupling model, may be due to the greater hardness of generated interaction matrices for discrete optimisation than continuous. The drastic difference between coupling models could be possibly mitigated with a better choice of $J_0$ or may be a signal of a better local minima escape mechanism of the \textit{relative} scheme. Nevertheless, the demonstration of the optimal performance of either of the proposed coupling methods is not the focus of this manuscript since both methods can be easily outperformed by standard heuristic algorithms. Instead, the achieved results clearly demonstrate a proof-of-principle for using polariton condensates, modelled with the mean-field approach equations (\ref{relative-rates}-\ref{nri}), as the XY-Ising computing machine. 

\textbf{Discussion}

\textit{Experimental Implementation.}
The spatially non-interacting condensates can be experimentally realised using lithographically etched micropillars or with trapped polariton condensates. The couplings are established remotely according to the elements of the coupling matrix $J_{ij}$. We envision two types of remote couplings. 
In the first scheme, the couplings are constructed by redirecting the emission of each condensate with either free-space optics or optical fibres to an SLM. At the SLM, the signal  from each node is multiplexed and redirected to other nodes with the desired coupling strength $J_{ij}$ allowing one, in principle, to create an all-to-all coupled network. Each  matrix of couplings $J$  can be programmed on the SLM in advance. 
 We refer to this implementation as \textit{all-optical} implementation. 
In the second approach,  the frequency and phase of the condensate emission are read out  and fed forward to an additional resonant excitation.  Such   resonant excitation  will have to be iteratively updated based on the  phase and energy of the emission until the polariton network synchronises. Consequently, the time-performance of the second scheme would be dependent on the operational frequency of the reading system and the SLM, which could be on the order of a few kHz  \cite{SLM2012}.


The comparable or better time-performance can be possibly achieved with the digital micro mirror devices which have a similar millisecond operational time-scale or with electro-optical modulators which can operate at nanosecond scale. We will refer to this implementation as \textit{hybrid-classical} implementation, since the condensate must first form to acquire a well-defined  phase that is read out and passed to other nodes. Note that in both implementations we consider symmetric interactions, i.e. $J_{ij} = J_{ji}$ for any two condensates in a network, though directional interactions can be readily constructed, e.g. by using an optical isolator.

In addition to two possible experimental implementations of the remote coupling control, we propose two kinds of couplings: \textit{absolute} and {\it relative}. The \textit{absolute} coupling scheme implies the exchange of equal amounts of photons (equal signals' intensities) between $i$-th and $j$-th nodes which guarantees that the occupation of the condensates pumped with equal intensities remains the same. In the {\it relative} coupling scheme, the condensates are coupled at the  rate defined by relative intensities of emission and, therefore, a further density adjustment is required \cite{KalininSciRep2018}. This adjustment is crucial for the operation of nonequilibrium condensates, lasers or DOPOs  as the density heterogeneity changes the values of the coupling strengths \cite{OurNJP2018}. Since the equilibration of densities will be done at the operation frequency of the SLM, the {\it relative} coupling model shares the same limitations as the \textit{hybrid-classical} implementation.


Thus, the \textit{absolute} coupling scheme with the \textit{all-optical} implementation may lead to a \textit{pure} polaritonic XY-Ising machine for optimising spin Hamiltonians since it doesn't require any external control: all couplings of a given spin Hamiltonian can be programmed on the SLM in advance. By approaching the condensation threshold from below, the polariton network will condense at one of the lowest energy states corresponding to a local or global minimum of the spin Hamiltonian. The term "pure" indicates that the system can operate at its own physical time-scale, i.e. picosecond scale for the polariton condensation. Among other pure physical simulators are the time-delay CIM \cite{time_delay_CIM2014} and the recently proposed pure molecular simulator \cite{Guzik_MolecularComputer2019}. The \textit{absolute} coupling scheme with the \textit{hybrid-classical} implementation as well as the {\it relative} coupling scheme with either of the proposed implementations would lead to the classical hybrid polariton simulators with an operational time limited by the frequency of the SLM. These approaches would be reminiscent of the CIM with a measurement feedback via FPGAs \cite{CIM2016} or hybrid molecular simulator \cite{Guzik_MolecularComputer2019}.
 
\textit{Polaritonic XY-Ising machine.}
In this work, we introduce a new approach for simulating discrete and continuous spin Hamiltonians, e.g. Ising and XY, with polariton networks. We propose two experimental implementations for realising remote phase locking of any two condensates in a micropillar array or in a lattice of trapped condensates with a potential to have fully-connected coupling matrices. The first scheme could possibly result in a pure optical polariton simulator in which the interactions are organised by redirecting the leaking photons from one condensate to another, therefore forming photonic feedback mechanism. The second leads to a hybrid-classical polariton simulator in which the interactions are realised with additional resonant injections. Both methods can be a viable option for building a real polaritonic XY-Ising machine. We verify the performance of the proposed machine by simulating polariton networks with the mean-field approach for two types of couplings between condensates: relative and absolute. Both methods clearly demonstrate the ability to optimise spin Hamiltonians of various sizes, up to 49 condensates, and various connectivities, up to 24 connections per element. Moreover, the possibility to simulate spin Hamiltonians with beyond nearest neighbour couplings is proposed for the first time in polaritonic networks. The real physical machine would benefit from a parallel-scanning through all phase configurations near the condensation threshold,  ultra-fast operational time-scale, high energy-efficiency with a milliwatt excitation power per condensate, and potential room-temperature operation.

\section*{Materials and Methods}
The numerical evolution of Eqs. (\ref{relative-rates}-\ref{nri}) is performed with the 4th-order Runge-Kutta time integration scheme and 4th order spatial finite difference scheme. The simulation parameters are $\eta = 0$, $g = 0.1$, $b_0 = 1$, $b_1 = 20$, $P({\bf r}, t) = \sum_i P_i \exp(-A \cdot |{\bf r - r}_i|^2)$ with $P_i = P_0 = 10$ for all micropillars in case of the absolute coupling scheme and dynamically adjusted $P_i$ to bring all the condensates to $\rho_{th} = 1$ in case of the relative coupling scheme for the XY Hamiltonian, $A = 5$, the distance between micropillars is $d = 2.4$,  $\gamma_{in} = 1$, $\gamma_{out} = 100$, $\alpha = 10$, $n_{SG} = 10$, the micropillar diameter is about $d_{micropillar} = 2$, $R \approx 0.25$. In addition, the following parameters are used to simulate the resonant pumping: $f_{\rm Res}({\bf r},t) = 50 ( \tanh (0.1 t - 3) + 1) \sum_{i=1}^N \exp(-25 |{\bf r - r}_i|)$ , where $t \in [0,T_{max}]$, $T_{max}$ is the time required to achieve a steady state.

\section{Additional Information}
The authors declare that they have no competing interests.

\section{Keywords}
Analogue computing machine, unconventional computing, exciton-polaritons, Ising Hamiltonian, XY Hamiltonian, polaritonic XY-Ising machine.


\end{document}